\documentclass[aps,preprint,groupedaddress,nofootinbib]{revtex4}
\usepackage{epsfig}
\usepackage{bm}

\newcommand{\ds}{\displaystyle}

\newcommand{\be}{\begin{equation}}
\newcommand{\ee}{\end{equation}}
\newcommand{\ba}{\begin{eqnarray}}
\newcommand{\ea}{\end{eqnarray}}
\newcommand{\etal}{\mbox{\it et al.}}

\begin{document}

\title{Deep exclusive electroproduction of $\pi^+$ at HERMES}
\author{Murat M. Kaskulov}
\email{murat.kaskulov@theo.physik.uni-giessen.de}
\author{Ulrich Mosel}
\affiliation{Institut f\"ur Theoretische Physik, Universit\"at Giessen,
             D-35392 Giessen, Germany}
\date{\today}

\begin{abstract}
Deeply virtual electroproduction of pion in exclusive reaction $p(e,e'\pi^+)n$
is studied using a two-component 
model which includes soft hadronic and hard
partonic reaction mechanisms. The results are compared with the experimental data measured
at HERMES in the deep inelastic region
for values of $Q^2>1$~GeV$^2$ and $W^2 > 10$~GeV$^2$. At forward angles the  $\pi^+$ cross
section is longitudinal and is 
dominated by exchange of Regge poles, with $\pi$ the
dominant trajectory. 
The off-forward region with $-t>1$~GeV$^2$ is transverse and
shows the dominance of partonic subprocesses. An implication of present
results for the future JLAB facilities are briefly discussed.
\end{abstract}
\maketitle


Exclusive electroproduction of pions in the reaction 
\be
e + N \to e' + \pi + N'
\ee
at high values of photon virtuality $Q^2$ and invariant mass $W$ of produced
hadronic final state
provides an interesting tool to study a space-times pattern of 
partonic interactions in deep inelastic scattering (DIS).
It may further reveal the partonic substructure of participating hadrons, 
correlating the longitudinal momentum fraction carried
by quarks to
transverse coordinates. This latter property of the hard exclusive 
reaction $N(e,e'\pi)N'$ follows from the QCD
factorization theorem which was proven for hard $Q^2\gg \Lambda_{\rm QCD}^2$
electroproduction of mesons by
longitudinal photons $\gamma^*_{\rm L}$~\cite{Collins:1996fb}. 
Predictions for the production by transverse virtual photons 
$\gamma^*_{\rm T}$ are
absent as no factorization theorem has been proven for such photons. However,
their contribution to the cross section is expected to be suppressed by at
least a power of $1/Q^2$.
On the other hand, above the resonance region $W^2>4$ GeV$^2$
the exclusive reaction $(\gamma^*,\pi^{\pm})$ with charged pions 
provides an important information concerning  
the electromagnetic form factor of the pion at momentum 
transfer $Q^2$ much bigger
than in the direct scattering of pions from atomic 
electrons~\cite{Sullivan:1970yq}.

Experimentally, the differential cross sections in the exclusive  reaction 
$p(\gamma^*,\pi^+)n$ has been measured above the resonance region 
at CEA~\cite{CEA}, Cornell~\cite{Cornell}, DESY~\cite{Desy}, and recently
at JLAB~\cite{Blok:2008jy}. At JLAB a
separation of cross sections into different transverse and longitudinal
components has been done.
The HERMES data at DESY~\cite{:2007an} largely extend the kinematic 
region to much higher values of $W$ toward the true DIS
region $Q^2\gg 1$~GeV$^2$ and 
much higher values of $-t$.
Theoretically, there is a long standing issue concerning the reaction 
mechanisms contributing
to the single pion $N(e,e'\pi)N'$ production at high
energies and photon virtualities. Just above the resonance region 
around the onset of the deep inelastic regime the models describing
the exclusive pion production $p(e,e'\pi^+)n$  in terms of hadronic 
degrees of freedom fail
to reproduce the large transverse cross section $\sigma_{\rm T}$ observed in this reaction.  
For instance, the hadron-exchange models which are generally considered to be a
guideline for the experimental analysis and extraction of the pion form
factor, underestimate grossly $\sigma_{\rm T}$ at the highest values of $Q^2$ 
measured at JLAB~\cite{Blok:2008jy}. While, the longitudinal cross section 
$\sigma_{\rm L}$ was supposed to be well understood
in terms of the pion quasi-elastic knockout
mechanism~\cite{Neudatchin:2004pu}. This is because of the pion pole at low $-t$. 
Even at smaller DESY~\cite{Desy} and 
much higher Cornell~\cite{Cornell}
values of $Q^2$ there is a disagreement between model
calculations  based on the hadron-exchange scenario and experimental data. 
Another interesting example is the neutral pion
production.
In $\pi^0$ photoproduction the cross section 
is well described at high energies by  exchange of Regge poles in the $t$-channel, 
with $\omega$ and $\rho$ the dominant
trajectories. This gives a natural explanation 
of the diffractive dip in the differential cross section at $-t\simeq 0.6$~GeV$^2$ 
provided a {\it wrong signature zero} 
is accounted for in the Regge amplitudes. 
However, already at low values of $Q^2$ the experimental data indicate
a sudden change in the reaction dynamics which 
washes out the diffractive dip.
The nature of this transition is not fully understood within 
Regge phenomenology.

A possible solution of
the $\sigma_{\rm T}$ problem at JLAB
was proposed in Ref.~\cite{Kaskulov:2008xc}. The approach followed there
is to complement the hadron-like interaction types in the $t$-channel,
which dominate in photoproduction and low $Q^2$ electroproduction, 
by a direct interaction of virtual photons with partons 
followed by string (quark) fragmentation into $\pi^+n$.   
Then $\sigma_{\rm T}$ 
can be readily explained and both $\sigma_{\rm L}$ and 
$\sigma_{\rm T}$ can be described from low up to high values of $Q^2$. 
In~\cite{Kaskulov:2008xc}
the reaction $p(e,e'\pi^+)n$ is treated as the exclusive limit, $z\to 1$, of 
the semi-inclusive  reaction $p(e,e'\pi^+)X$ in DIS in the spirit of
the exclusive-inclusive connection~\cite{Bjorken:1973gc}.
A suggestion concerning the partonic  
contribution to the exclusive  reaction $N(e,e'\pi)N'$ 
follows the arguments in~\cite{Nachtmann:1976be}
where it has been shown that the typical exclusive photoproduction 
mechanisms involving a
peripheral quark-antiquark pair in the proton wave function, 
the $t$-channel
meson-exchange processes, should be unimportant in the
transverse response $\sigma_{\rm T}$ already around
$Q^2 \gtrsim 1$~GeV$^2$ and play no role in the true deep inelastic region.

\begin{figure}[h]
\begin{center}
\includegraphics[clip=true,width=0.48\columnwidth,angle=0.] 
{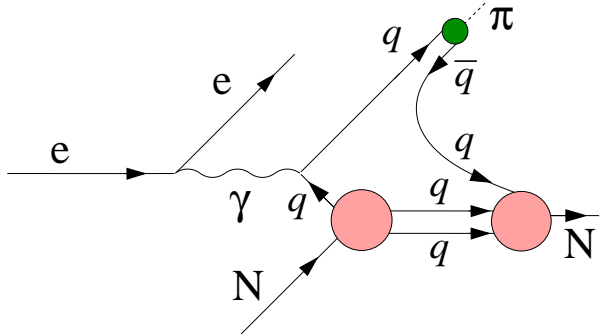}
\caption{\label{FigDIS} 
\small (Color online) A schematic representation of the deeply virtual partonic part of the $\pi$-
electroproduction mechanism. 
See the text for the details.
\vspace{-0.0cm} } 
\end{center}
\end{figure}

In this brief report we apply the two-component hadron-parton model proposed in 
Ref.~\cite{Kaskulov:2008xc} for the description of recent HERMES data
in order to check its validity in a different kinematic regime. We use the
same model parameters as in~\cite{Kaskulov:2008xc}.
In the model of~\cite{Kaskulov:2008xc}  the hadron exchange part is
described by exchange of Regge trajectories. In the reaction $p(\gamma^*,\pi^+)n$ we
take into account the exchange of $\pi(140)$ and 
$\rho(770)$-meson Regge trajectories. The former one includes the
electric part of the nucleon-pole contribution to conserve the charge of the
system~\cite{Vanderhaeghen:1997ts}. The partonic part is shown schematically in Figure~\ref{FigDIS}. It is
a DIS like electroproduction mechanism where the quark knockout
reaction $\gamma^* q \to q$ is followed by the 
fragmentation process of the Lund type~\cite{Andersson:1983ia}.  
We  refer to~\cite{Kaskulov:2008xc} for further details. 
We recall that at JLAB kinematics the transverse cross section is
dominated by the mechanism described in Figure~\ref{FigDIS}. 
As a result the transverse cross section is large and at forward angles is comparable
with the longitudinal cross section. As we shall see at HERMES, where the value of $W$ is much larger, the
transverse cross section gets much smaller at forward angles as compared to
JLAB and the cross section is
dominated by exchange of Regge trajectories. However, the situation is
different in the off-forward region. Because of the exponential fall-off of
Regge contributions as a function of $-t$ the
meson-exchange processes are negligible in the region of $-t>1$~GeV$^2$. In this
region the interaction of virtual photon with partons
gives the dominant contribution.

\begin{figure}[b]
\begin{center}
\includegraphics[clip=true,width=0.8\columnwidth,angle=0.] 
{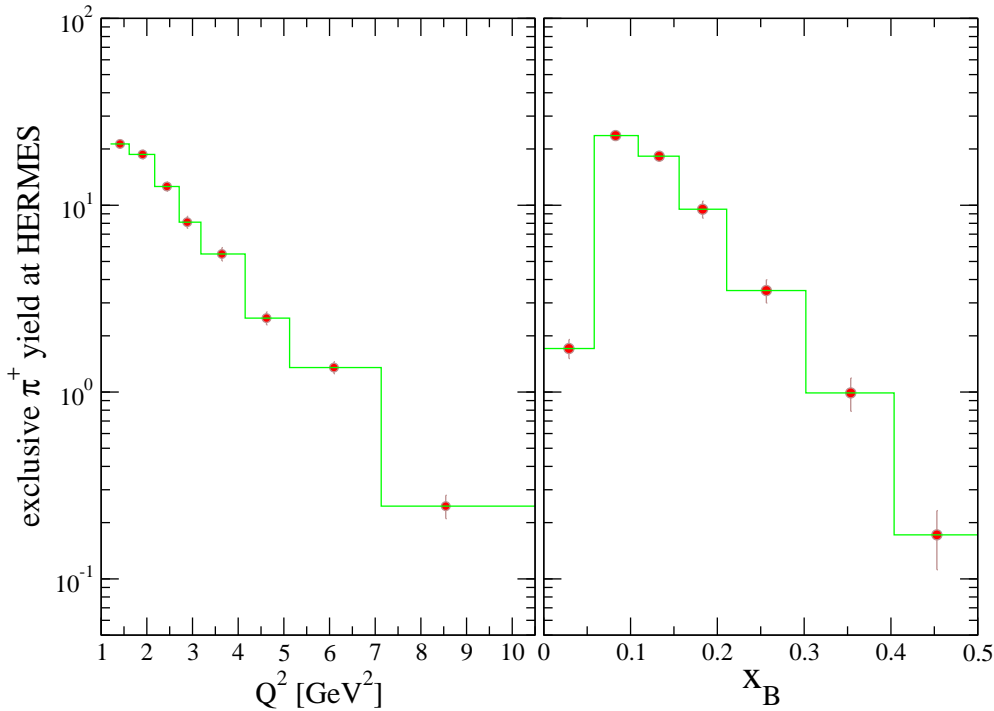}
\caption{\label{HermesAcceptance} 
\small (Color online) Distribution of exclusive $\pi^+$ events within the HERMES acceptance
as a function of $Q^2$ and Bjorken scaling variable $x_B$.
\vspace{-0.0cm}  
}
\end{center}
\end{figure}

At first, a brief discussion concerning some features of HERMES data is in
order.
The differential cross section in the exclusive reaction $p(e,e'\pi^+)n$ 
can be written in the following form
\begin{eqnarray}
\label{dsdte}
\frac{d\sigma}{dQ^2d\nu dt} &=&
\frac{\pi \Phi}{E_e(E_e-\nu)}
\left[ 
           \frac{d\sigma_{\rm T}}{dt} 
+ \varepsilon \frac{d\sigma_{\rm L}}{dt} \right],
\end{eqnarray}
where the virtual photon flux is given by~\cite{Hand:1963bb}
\begin{equation}
\Phi = 
\frac{\alpha}{2\pi^2} \frac{E_e-\nu}{E_e} 
\frac{\mathcal{K}}{Q^2}
\frac{1}{1-\varepsilon},
\end{equation}
with $\alpha \simeq 1/137$, $\mathcal{K}=\nu(1-x_{B})$, $\nu=E_e-E_e'$ and 
\begin{equation}
\varepsilon = 
              \frac{1}{\ds 1+2 \frac{\nu^2+Q^2}{4E_e(E_e-\nu)-Q^2}}
\end{equation}
is the ratio of longitudinal to transverse polarization of the virtual
photon and other notations are obvious.
At HERMES the kinematic requirement $Q^2>1$~GeV$^2$ has been imposed on the
scattered lepton in order to select the hard scattering regime. The
resulting kinematic range is $1<Q^2<11$~GeV$^2$ and $0.02<x_B<0.55$ for the
Bjorken variable. 
The measured
cross sections are integrated over the azimuthal angle $\varphi$ and a
separation of the transverse and longitudinal parts was not feasible.
Furthermore, the exclusive data were obtained from $\pi^+$ semi-inclusive DIS data 
by using a model dependent subtraction procedure. 
With $E_e=27.6$~GeV beam energy at DESY the ratio of 
longitudinal to transverse polarization of the virtual photon $\varepsilon$
is high $\varepsilon \simeq 0.95$. Contrary to JLAB where $\varepsilon$ is smaller 
the longitudinal cross section at HERMES 
enters the differential cross section with practically its full strength.
The value of $s=W^2$ was required to be higher than
10~GeV$^2$. 
To make a proper comparison with data the theoretical cross 
sections have to be corrected for the bin size effect to account
for the $Q^2$ and $x_B$ dependence of the exclusive $\pi^+$ yield  within the 
HERMES acceptance.  The corresponding experimental distributions of $\pi^+$  
as a function of $Q^2$ (left) and $x_B$ (right) are shown in
Figure~\ref{HermesAcceptance} and are used to generate the virtual photon flux in
the Monte Carlo procedure to fold the theoretical cross sections.

\begin{figure*}[t]
\begin{center}
\includegraphics[clip=true,width=1\columnwidth,angle=0.] 
{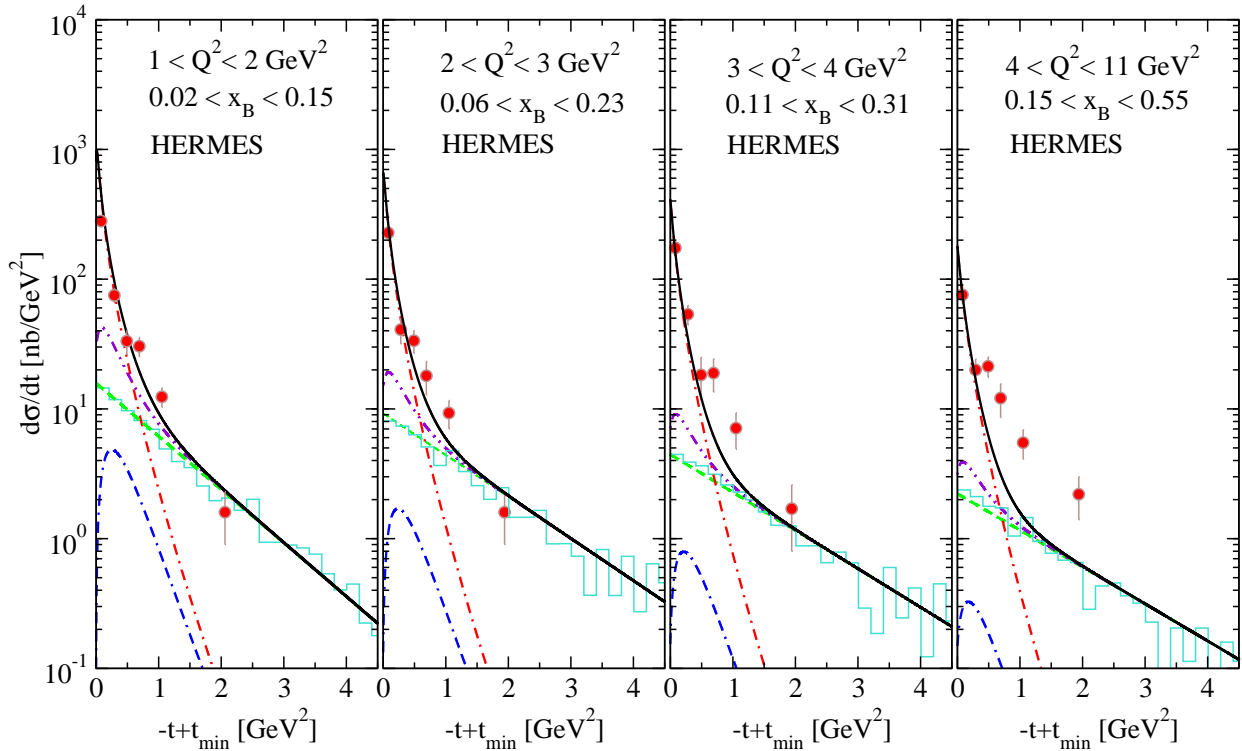}
\caption{\label{Figure1}
\small (Color online) The differential cross section 
$d\sigma/dt = d\sigma_{\rm T}/dt + \epsilon d\sigma_{\rm
L}/dt$ of the exclusive reaction $p(\gamma^*,\pi^+)n$ in the kinematics of HERMES
experiment. 
The solid curves are the model results.
The dash-dotted curves correspond to the exchange of $\pi$-Regge trajectory and
dash-dash-dotted curves to the exchange of $\rho$-Regge trajectory. 
The histograms and the dashed curves which just fit the histograms describe 
the partonic contributions in line of the DIS mechanism shown in
Figure~\ref{FigDIS}. The dot-dot-dashed curves are the contribution 
of the transverse cross section $d\sigma_{\rm T}/dt$ to the total unseparated response 
$d\sigma/dt$.
\vspace{-0.0cm} 
}
\end{center}
\end{figure*}

Our results for the photon-nucleon differential cross section 
$d\sigma=d\sigma_{\rm T}+\varepsilon d\sigma_{\rm L}$ in the reaction
$p(\gamma^*,\pi^+)n$ at HERMES are shown in Figure~\ref{Figure1}.
In Figure~\ref{Figure1} instead of $t$, the quantity $-t+t_{min}$ is used, where $-t_{min}$
denotes the minimum value of $-t$ for a given $Q^2$ and $x_{B}$.
Different panels in Figure ~\ref{Figure1} correspond to different $Q^2$ and $x_B$ bins.
The model results which include both the hadron-exchange and partonic DIS contributions
are shown by the solid curves. We obtain satisfactory
agreement with data for first three $(Q^2,x_B)$ bins and underestimate
the cross section for the last bin with the highest values of $Q^2$.
Although with the model parameters at hand a perfect description of data can be
achieved, we do not attempt to fit the data  and use the same model parameters
tuned to JLAB data. We recall again that the experimental data themselves include 
a model dependent background subtraction uncertainties.

Concerning different contributions to the cross section, in Figure~\ref{Figure1} 
the dash-dotted curves correspond to the exchange of $\pi$-Regge trajectory and
dash-dash-dotted curves to the exchange of $\rho$-Regge trajectory.
All the trajectories
used here are linear. The histograms and dashed curves which just 
fit the histograms describe 
the hard partonic contributions in line of the DIS mechanism shown in
Figure~\ref{FigDIS}, 
see for the details Ref.~\cite{Kaskulov:2008xc}.
As one can see, at forward angles the $\pi$-exchange dominates the
differential cross sections. 
The steep fall of $d\sigma/dt$ away from forward
angles comes entirely from the rapidly decreasing $\pi$-exchange amplitude.
The $\pi$-exchange contributes mainly to the longitudinal response
$\sigma_{\rm L}$ and at low momentum
transfer $-t$ the variation of the cross section with $Q^2$ falls down as
the electromagnetic form factor of the pion
$\sigma_{\rm L} \propto (F_{\gamma\pi\pi} (Q^2))^2$ 
\begin{equation}
\label{PiFF}
F_{\gamma\pi\pi} (Q^2) =
{(1+Q^2/\Lambda_{\gamma\pi\pi}^2)^{-1}}.
\end{equation} 
The value of the cut-off in Eq.~(\ref{PiFF}) used in the calculations 
is $\Lambda_{\gamma\pi\pi}^2=0.52~$GeV$^2$. This is an
optimal value needed to describe the JLAB data in the present model 
for values of $Q^2 \gtrsim
1.5$~GeV$^2$. The magnitude of the cross section and slope of solid curves 
and data at very forward angles 
are consistent with the Regge behavior
\be
d\sigma/dt \sim
e^{2 \alpha_{\pi}'\ln(s/s_0)(t-m^2_{\pi}) - 2\ln(s/s_0)},
\ee
where $\alpha'_{\pi}=0.7$~GeV$^{-2}$ and $s_0=1$~GeV$^2$. 
The contribution of the 
$\rho$-exchange is marginal
in $\sigma_{\rm L}$ and $\sigma_{\rm T}$.  

At large $-t$ the model cross section
points mainly toward the direct coupling of the virtual photons to
partons. Indeed, this is rather natural, since with 
increasing $-t$ at fixed $Q^2$  smaller distances
can accordingly be accessed. This is opposed to $t$-channel meson-exchange 
processes which involve peripheral production of $\pi^+$ and therefore 
large distances from the origin.
At forward angles the DIS contribution
(histograms in Figure~\ref{FigDIS}) to the 
cross section is by two orders of magnitude smaller
than
the meson-exchange contributions. However, the slope of the cross section
in this case is also much smaller. 
Note that, in the model of Ref.~\cite{Kaskulov:2008xc} the partonic part
is transverse and its behavior (slope) at forward angles is driven by the intrinsic transverse
momentum distribution of partons $\langle k_{\rm T}^2\rangle$. From the fit to
the histograms (dashed curves)
we observe a slight decrease of the slope as a function 
of $Q^2$ but it is essentially the same as
at JLAB energies. 
The partonic DIS part of the cross section 
shows a rather weak $Q^2$ dependence  
and has approximately the same order of magnitude for all $Q^2$ bins.
The dot-dot-dashed curves in Figure~\ref{FigDIS} are the contribution 
of the transverse cross section $d\sigma_{\rm T}/dt$ which includes both Regge
and hard partonic mechanisms to the total unseparated differential cross section
$d\sigma/dt$.

\begin{figure}[t]
\begin{center}
\includegraphics[clip=true,width=0.85\columnwidth,angle=0.] 
{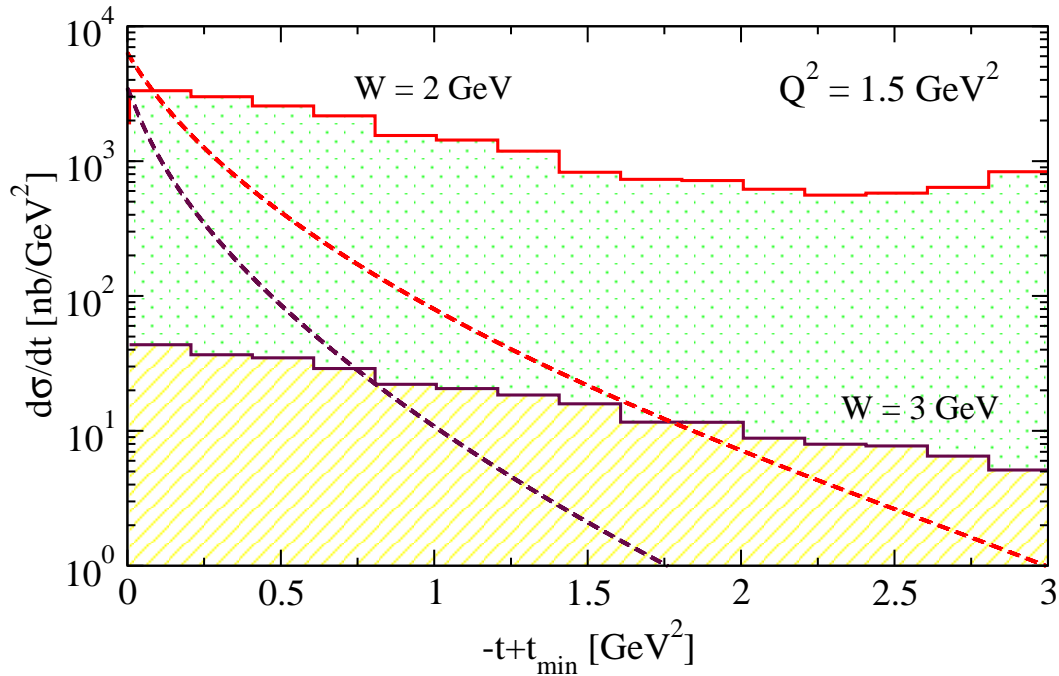}
\caption{\label{Figure4} 
\small (Color online) The transverse DIS (histograms) and longitudinal Regge-exchange (dashed
curves) differential cross sections in the exclusive reaction $p(\gamma^*,\pi^+)n$ 
for the fixed value of $Q^2=1.5$~GeV$^2$ and for different values of 
$W=2$~GeV (top) and $W=3$~GeV (bottom).  \vspace{-0.0cm}  
}
\end{center}
\end{figure}

The partonic interpretation of the large $-t$ region proposed here
is similar to the interpretation by~\cite{Laget:2004qu} of HERMES data.
In the latter work the pion form factor has been modified to account
for different space-time patterns of hard interaction processes. However, there is
an important difference. By modifying the off-mass-shell behavior of 
$F_{\gamma\pi\pi}(Q^2)$ 
at large $-t$, see Eq.~(\ref{PiFF}) with 
$\Lambda_{\gamma\pi\pi}\to \Lambda_{\gamma\pi\pi}(t,Q^2)$,  
one essentially modifies the dominant longitudinal component and absorbs the
partonic contributions in $\sigma_{\rm L}$. Our interpretation
relies on DIS like $\gamma^* q \to q$ interaction which shows up in the 
transverse response $\sigma_{\rm T}$ as in any high $(Q^2,W)$ DIS processes.

The considerable reduction of the transverse exclusive cross section 
as a function of $W$ observed here 
has important consequences for the future JLAB facilities.
For instance, in Figure~\ref{Figure4} we show the longitudinal hadronic 
and transverse partonic cross sections at values of $Q^2=1.5$~GeV$^2$ and 
for two different values of $W=2$~GeV and
3~GeV. The transverse cross section (top histogram) at present JLAB energies 
($W=2$~GeV) is large in agreement with data and at forward directions 
is comparable in order of magnitude 
with the longitudinal cross section (dashed curve). 
An increase of $W$ just from 2~GeV to 3~GeV reduces the transverse DIS cross
section by about two orders of magnitude. However, the longitudinal cross
section is barely changed. 
This may support QCD based calculations which rely on small
transverse components in an extraction of nucleon's GPD.  
This example also shows that at JLAB with 12 GeV
electron beam an extraction 
of the pion charge form factor from forward exclusive electroproduction 
data will not be
contaminated by the large transverse non-pole background contributions. Furthermore, 
at these energies the theoretical analysis and proof of the color transparency signal 
observed in semi-exclusive $(e,e'\pi^+)$ off nuclei will be
considerably simplified~\cite{:2007gqa,Kaskulov:2008ej}. 

In summary, we applied the model of Ref.~\cite{Kaskulov:2008xc} to recent 
HERMES data on exclusive
$\pi^+$ electroproduction in the deep inelastic region. The data were obtained
at much higher values of $Q^2$ and $W$ as compared to the previous data from
JLAB. We find that the transverse cross section, which is large at JLAB, gets much
smaller at HERMES. However, with increasing $-t$ the role of partonic
DIS mechanisms becomes 
more pronounced. We also find that the forward production of $\pi^+$ at HERMES 
is dominated by exchange of Regge poles, with $\pi$ being the dominance trajectory. 
On the contrary, the off-forward region with
$-t>1$~GeV$^2$ is dominated by partonic interactions describing the
direct coupling of virtual photons to constituents of the nucleon.
For future JLAB energies the model predicts a tiny
transverse component which might be important for the GPD interpretation
of $p(\gamma^*,\pi^+)n$
as well as in the extraction 
of the pion form factor from the forward electroproduction data.

We are grateful to Dr. A.~Airapetian for helpful communications and
for making the actual $(Q^2,x_B,-t)$ values of the data available to us. 

This work was supported by BMBF.


\begin{thebibliography}{100}

\bibitem{Collins:1996fb}
  J.~C.~Collins, L.~Frankfurt and M.~Strikman,
  Phys.\ Rev.\  D {\bf 56}, 2982 (1997).

\bibitem{Sullivan:1970yq}
  J.~D.~Sullivan,
  Phys.\ Lett.\  B {\bf 33}, 179 (1970).




\bibitem{CEA}
C.~N.~Brown \etal,
 Phys.\ Rev.\  D {\bf 8}, 92 (1973).
\bibitem{Cornell}
C.~J.~Bebek \etal,
Phys.\ Rev.\  D {\bf 17}, 1693 (1978).
\bibitem{Desy}
P.~Brauel \etal,
Phys.\ Lett.\  B {\bf 69}, 253 (1977). 


\bibitem{Blok:2008jy}
  H.~P.~Blok {\it et al.},
  Phys.\ Rev.\  C {\bf 78}, 045202 (2008).




\bibitem{:2007an}
  A.~Airapetian {\it et al.}, 
  Phys.\ Lett.\  B {\bf 659}, 486 (2008).



\bibitem{Neudatchin:2004pu}
  V.~G.~Neudatchin {\it et al.}, 
  Nucl.\ Phys.\  A {\bf 739}, 124 (2004).



\bibitem{Kaskulov:2008xc}
  M.~M.~Kaskulov, K.~Gallmeister and U.~Mosel,
  Phys.\ Rev.\  D {\bf 78}, 114022 (2008).




\bibitem{Bjorken:1973gc}
  J.~D.~Bjorken and J.~B.~Kogut,
  Phys. Rev. D {\bf 8}, 1341 (1973).



\bibitem{Nachtmann:1976be}
  O.~Nachtmann,
  Nucl.\ Phys.\  B {\bf 115}, 61 (1976).


\bibitem{Vanderhaeghen:1997ts}
  M.~Vanderhaeghen, M.~Guidal and J.~M.~Laget,
  Phys.\ Rev.\  C {\bf 57}, 1454 (1998).


\bibitem{Andersson:1983ia}
  B.~Andersson {\it et al.}, 
  Phys.\ Rept.\  {\bf 97}, 31 (1983).

\bibitem{Hand:1963bb}
  L.~N.~Hand,
  Phys.\ Rev.\  {\bf 129}, 1834 (1963).



\bibitem{Laget:2004qu}
  J.~M.~Laget,
  Phys.\ Rev.\  D {\bf 70}, 054023 (2004).


\bibitem{:2007gqa}
  B.~Clasie {\it et al.},
  Phys.\ Rev.\ Lett.\  {\bf 99}, 242502 (2007).

\bibitem{Kaskulov:2008ej}
  M.~M.~Kaskulov, K.~Gallmeister and U.~Mosel,
  Phys.\ Rev.\  C {\bf 79}, 015207 (2009).





\end{thebibliography}
\end{document}